\documentclass[notitlepage,author-numerical,showkeys,showpacs,11pt]{revtex4-1}
\usepackage{amsmath,amsthm,graphicx,amssymb,natbib}
\usepackage{float}
\usepackage{subfig}
\usepackage{bm}
\usepackage{tikz}
\usepackage{epstopdf}
\begin{document}
\setcitestyle{super}
\title{Collapse and Revival Oscillation in Double Jaynes-Cummings Model}
\author{Surajit Sen$^{\S,\star}$}
\email[Corresponding author:\quad]{ssen55@yahoo.com}
\author{Tushar Kanti Dey$^{\S,\star}$}
\email[]{tkdey54@gmail.com}
\author{Mihir Ranjan Nath$^{\S,\star}$}
\email[]{mrnath_95@rediffmail.com}
\affiliation{$^{\S}$ Physics Department, Guru Charan College, Silchar 788004, India}
\affiliation{$^{\star}$ Centre of Advanced Studies and Innovation Lab, 18/27 Kali Mohan Road, Tarapur, Silchar 788003, India}

\begin{abstract}
We develop a systematic method of solving two noninteracting Jaynes-Cummings models by using the dressed state formalism in Hilbert space $\mathcal{H}_{AB}^{(2\otimes2)}$. It is shown that such model, called Double Jaynes-Cummings model (D-JCM), can be exactly solved if we take the initial bare state as the linear superposition of two Bell states. The collapse and revival oscillation, which is the standard trait of typical Jaynes-Cummings model, can be recovered if we make measurement at each local sites. Some consequence of the entanglement-induced dressing is discussed.
\end{abstract}
\maketitle
\section{\label{1x} Introduction}
\par
To address some intrinsic riddle of the quantum mechanics of separated bodies and their separability, in 1935 Schr{\"{o}}dinger introduced the notion of entanglement which has become an integral part of quantum information science \cite{schrodinger1935,schrodinger1936}. Inspite of its counter-intuitive origin, today entanglement is believed to be the primary ingredient of several charted and uncharted issues of quantum teleportation \cite{bb1984}, quantum cryptography \cite{rsa1978,ekert1991} and quantum computer \cite{nielsen2000}, the core parts of much-touted quantum technology. The key feature of entanglement is that, when a single system is parted away into apparently two isolated subsystems, then any measurement in one subsystem influences other, no matter whatever be the distance between them. Such nonlocal correlation, which is often referred as the entanglement or quantum correlation, is quite different from the local gauge theory where the causal interaction between two subsystems is mediated by so-called `gauge particle'. In a pioneering experiment, Aspect and his coworkers \cite{aspect1982} conclusively verified the violation of the Bell-CHSH inequality \cite{bell1964,chsh1969} which uphold the result of quantum mechanics invalidating the hidden variable theory advocated by Einstein-Podolsky-Rosen (EPR) \cite{epr1935}. In the subsequent years, the `spooky' tag of the quantum correlation was completely removed by several loophole free precision experiments \cite{hensen2015,shalm2015}. Apart from the epistemological issues like local realism and measurement, other core problem is the proper quantification of the quantum correlation and in this regard, concurrence \cite{wootters1997,wootters1998,wootters2006}, I-concurrence \cite{rungta2001}, quantum discord \cite{ollivier2001}, entanglement witness \cite{lewenstein2001,guhne2009} etc have been proposed as the possible measure of the quantum correlation with different physical attributes.
\par
Simplest example of an entangled system is the bipartite system which is formed by two nonlocal qubits called two-qubit system. Conventionally, a pair of photon, each with two state of polarization produced in the parametric down conversion process, is an efficient representative of the bipartite system. In the recent past, Eberly and his coworkers argued that a pair of two non-interacting two-level systems produced during the atomic or molecular dissociation, can also be regarded as the bipartite system with several extra features \cite{eberly2006,eberly2007,eberly2009}. Their study reveals that the intrinsic Rabi oscillation of the two-level system is manifested through an effective oscillation of the \textit{concurrence} \cite{wootters1997,wootters1998} which also exhibits \textit{entanglement sudden death} due to decoherence \cite{eberly2006,eberly2009}. Soon after their proposal, there is an upsurge of studying various quantum-optical effect in presence of entanglement in various routes \cite{eberly2006,eberly2007,eberly2008,malinovsky2006,saha2010,jarvis2010,jinshi2010,bahari2018,yan2014,Sainz2007,
Han2010}, however, the exact solution of Double Jaynes-Cummings model (D-JCM) using dressed state formalism is still not available.
\par
There exists a wide class of the coherent phenomena which arises due to the multiple level structure of a quantum optical system. Among them, the four-level system defined in Hilbert space $\mathcal{H}^{(4)}$, is related with the phenomena like four-level EIT effect \cite{joshi2003}, pulse propagation in coherently prepared four-level system \cite{paspalakis1999,paspalakis2002}, qubit-induced micro-switching \cite{ham2000,harris1998}, Rabi oscillation in the equidistant four-level system \cite{nath2008} etc. Recently, in a series of work we have developed a dressed state formalism to solve the three \cite{sen2012} and four-level system \cite{nath2008,sen2014} and studied their Rabi oscillation and other possible properties. To generalize that approach, in the present paper we extend the dressed state formalism to solve the D-JCM in the Bell basis. We explicitly show that the projection measurement of the entangled system defined in $\mathcal{H}_{AB}^{(2\otimes2)}$,  which amounts to taking the partial trace over the subsystem, precisely gives the population inversion and collapse and revival oscillation of typical JCM at each local sites.
\par
To achieve our objective, remaining Sections of the paper are organized as follows: In Section II, we introduce the D-JCM in a requisite form and define the appropriate bare state in the Bell basis. In Section III we develop the dressed state formalism to solve the model. The population inversion scenario and the collapse and revival oscillation of the D-JCM are studied in Section IV  and their  possible consequences are discussed. We conclude by summarizing essential results of our investigation and discuss possible outlook.

\section{The double Jaynes-Cummings model}
\subsection{\label{2a} The model:}
\par
In the rotating wave approximation, the Hamiltonian of the Jaynes-Cummings model in Hilbert space $\mathcal{H}_i^{(2)}$ is given by \cite{jcm1963,barnett1997},
\begin{equation}\label{one}
H_{i}=\omega_ia_i^\dagger a_i+\frac{\Delta}{2} \sigma^z_{i}+g_i(\sigma^{+}_ia_i+h.c.),
\end{equation}
where, $\sigma_i^{+}$ ($\sigma_i^{-}$)and $a^{\dagger}_i$ ($a_i$) be the `atom' and `field' creation (annihilation) operators with $g_i$ be the coupling parameter at local $i$-th sites ($i=$A (Alice), B (Bob)) and $\Delta$ be the detuning, respectively.
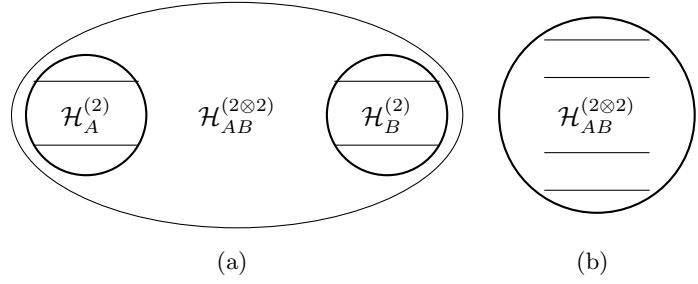
\begin{figure}\centering
\subfloat[]{
\begin{tikzpicture}[baseline]
\draw [] (5.7,1.95) -- (4.3,1.95) (5.7,1.1) -- (4.3,1.1);
\draw [] (1.7,1.95) -- (0.3,1.95) (1.7,1.1) -- (0.3,1.1);
\draw [thick] (1,1.5) circle [radius=.8] node[] {$\mathcal{H}_{A}^{(2)}$};
\draw [thick] (5,1.5) circle [radius=.8] node[] {$\mathcal{H}_{B}^{(2)}$};
\tikz \draw (3,1.5) ellipse (3cm and 1.5cm) node[] {$\mathcal{H}_{AB}^{(2\otimes2)}$};
\end{tikzpicture}}\quad
\subfloat[]{
\begin{tikzpicture}[baseline]
\draw [thick] (1,1.5) circle [radius=1.3] node[] {$\mathcal{H}_{AB}^{(2\otimes2)}$};
\draw [] (1.7,2.5) -- (0.3,2.5) (1.7,0.5) -- (0.3,0.5);
\draw [] (1.7,2.0) -- (0.3,2.0) (1.7,1) -- (0.3,1);
\end{tikzpicture}}
\caption{Two two-level systems in product Hilbert space $\mathcal{H}_{AB}^{(2\otimes2)}$ shown in (a) may be viewed as an effective four-level system (b).}
\end{figure}
Fig.1 shows the schematic diagram of two two-level systems entangled together to form an effective four-level system Hilbert space $\mathcal{H}_{AB}^{(2\otimes2)}$. Thus the Hamiltonian of the Double Jaynes-Cummings system, which is indeed a bipartite system formed by two nonlocal two-level quantized systems, is given by,
\begin{equation}
\begin{split}
\hat{H}_{AB}&=\omega_A\mathbb{I}_A\otimes\mathbb{I}_Ba_A^\dagger a_A+\omega_B\mathbb{I}_A\otimes\mathbb{I}_Ba_B^\dagger a_B +
\frac{\Delta}{2} \mathbb{I}_A\otimes\sigma^z_B + \frac{\Delta}{2} \sigma^z_A\otimes\mathbb{I}_B   \\
&+g_{B}\mathbb{I}_A\otimes\sigma^+_Ba_B +g_{B}\mathbb{I}_A\otimes\sigma^-_Ba_B^\dagger + g_{A}a_A\sigma^+_A\otimes\mathbb{I}_B+g_{A}a_A^\dagger\sigma^-_A\otimes\mathbb{I}_B.
\end{split}
\end{equation}
At zero detuning $\Delta=0$, the off-diagonal contribution of the Hamiltonian can be equivalently expressed in a $4\times4$ matrix,
\begin{equation}\label{three}
\hat{H}_{AB}^I=\left(
  \begin{array}{cccc}
    0 & g_{B}a_{B} & g_{A}a_{A} & 0 \\
    g_{B}a_{B}^\dag & 0 & 0 & g_{A}a_{A} \\
    g_{A}a_{A}^\dag & 0 & 0 & g_{B}a_{B} \\
    0 & g_{A}a^\dag_{A} & g_{B}a^\dag_{B} & 0
  \end{array}
\right),
\end{equation}
which we proceed to solve using the dressed state formalism  \cite{nath2008,sen2014} for suitable bare state.

\subsection{\label{4a} Bare state wave function:}
\par
A linear combination of two-qubit states constitutes the Bell states,
\begin{subequations}
\begin{eqnarray} \label{four}
\mid \Phi^{+}\rangle_{AB}&=\frac{1}{\sqrt{2}}(\mid 0_{A}1_{B}\rangle{+}\mid 1_{A}0_{B}\rangle), \\
\mid \Psi^{+}\rangle_{AB}&=\frac{1}{\sqrt{2}}(\mid 0_{A}0_{B}\rangle{+}\mid 1_{A}1_{B}\rangle), \\
\mid \Psi^{-}\rangle_{AB}&=\frac{1}{\sqrt{2}}(\mid 0_{A}0_{B}\rangle{-}\mid 1_{A}1_{B}\rangle), \\
\mid \Phi^{-}\rangle_{AB}&=\frac{1}{\sqrt{2}}(\mid 0_{A}1_{B}\rangle{-}\mid 1_{A}0_{B}\rangle),
\end{eqnarray}
\end{subequations}
respectively, and most general bare state wave function of a semiclassical D-JCM is given by,
\begin{equation}\label{five}
\begin{split}
\mid \psi_{AB}^{}(0)\rangle &= c_{00}(0)\mid \Phi^+\rangle_{AB} +c_{01}(0)\mid \Psi^+\rangle_{AB} \\
&+c_{10}(0)\mid \Phi^-\rangle_{AB} +c_{11}(0)\mid \Psi^-\rangle_{AB},
\end{split}
\end{equation}
where $c_{ij}$ ($i, j=0, 1$) are the normalized amplitudes. Taking the cavity mode at each local site quantized, the bare state wave function of the quantized D-JCM can be written as,
\begin{equation}\label{six}
\begin{split}
\mid \psi_{AB}(0) \rangle =\sum\limits_{n_A,n_B=0}^{\infty} &\{c_{00}(0)\mid \Phi^{+};{n_A,n_B}\rangle
+c_{01}(0)\mid \Psi^{+};{n_A,n_B}\rangle \\
&+c_{10}(0)\mid \Phi^{-};{n_A,n_B}\rangle +c_{11}(0)\mid \Psi^{-};{n_A,n_B}\rangle\},
\end{split}
\end{equation}
where, $n_A$ and $n_B$ be the field modes at local sites $A$ and $B$ with the Bell-Fock basis given by,
\begin{subequations}\label{seven}
\begin{eqnarray}
\mid \Phi^{\pm};n_A,n_B\rangle &=\frac{1}{\sqrt{2}}(\mid 0_{A}1_{B};n_A+1,n_B \rangle {\pm}\mid 1_{A}0_{B};n_A,n_B+1\rangle), \\
\mid \Psi^{\pm};n_A,n_B\rangle &=\frac{1}{\sqrt{2}}(\mid 0_{A}0_{B};n_A+1,n_B +1\rangle{\pm}\mid 1_{A}1_{B};n_A,n_B\rangle),
\end{eqnarray}
\end{subequations}
respectively. It is worth noting here that, none of these four Bell-Fock states in Eq.(7) separately constitutes a solution of Schrodinger equation, but a pair of them does. In particular, we note that out of $\frac{4!}{(4-2)!2!}=6$ independent Bell pairs, only following linear combinations, namely, $\{\mid \Phi^{+}\rangle,\mid \Psi^{+}\rangle \}$ and $\{\mid \Phi^{-}\rangle,\mid \Psi^{-}\rangle \}$, are exactly solvable. Thus, there exists two distinct situations for which the interaction Hamiltonian in Eq.(3) is exactly solvable:\\

Scenario-I: When the initial bare state is formed by superposing the Bell pair, $\mid \Phi^{+}\rangle$ and $\mid \Psi^{+}\rangle$:\\

By substituting Eq.(7) into Eq.(6) and taking $c_{10}(0)=0, c_{11}(0)=0$, we obtain following bare state wave function,
\begin{equation}\label{eight}
\begin{split}
\mid \psi_{AB}^{I}(0) \rangle = \sum\limits_{n_A,n_B=0}^{\infty}& c_{01}(0)\{\mid 0_A0_B;{n_A+1,n_B+1}\rangle
+\mid 1_A1_B;{n_A,n_B}\rangle \} \\
&+ c_{00}(0)\{\mid 0_A1_B;{n_A+1,n_B}\rangle +\mid 1_A0_B;{n_A,n_B+1}\rangle \},
\end{split}
\end{equation}
which can be equivalently written as,
\begin{equation}\label{nine}
\mid \psi_{AB}^{I}(0) \rangle =\sum\limits_{n_A,n_B=0}^{\infty}\left(
  \begin{array}{c}
   c_{00}(0) \mid {n_A+1,n_B+1}\rangle\\
   c_{01}(0) \mid {n_A+1,n_B}\rangle\\
   c_{01}(0) \mid {n_A,n_B+1}\rangle\\
   c_{00}(0) \mid {n_A,n_B}\rangle
  \end{array}
\right).
\end{equation}
\par
Scenario-II: When the initial bare state is formed by the Bell pair $\mid \Phi^{-}\rangle$ and $\mid \Psi^{-}\rangle $:\\
Proceeding similar way and taking $c_{00}(0)=0, c_{01}(0)=0$ we obtain,
\begin{equation}\label{ten}
\begin{split}
\mid \psi_{AB}^{II}(0) \rangle = &\sum\limits_{n_A,n_B=0}^{\infty}c_{11}(0)\{\mid 0_A0_B;{n_A+1,n_B+1}\rangle -\mid 1_A1_B;{n_A,n_B}\rangle \} \\
& +c_{10}(0)\{\mid 0_A1_B;{n_A+1,n_B}\rangle -\mid 1_A0_B;{n_A,n_B+1}\rangle\},
\end{split}
\end{equation}
which can be also expressed as,
\begin{equation}\label{eleven}
\mid \psi_{AB}^{II}(0) \rangle =\sum\limits_{n_A,n_B=0}^{\infty}\left(
  \begin{array}{c}
   c_{10}(0) \mid {n_A+1,n_B+1}\rangle\\
   c_{11}(0) \mid {n_A+1,n_B}\rangle\\
   -c_{11}(0) \mid {n_A,n_B+1}\rangle\\
   -c_{10}(0) \mid {n_A,n_B}\rangle\\
  \end{array}
\right).
\end{equation}
After developing the requisite model and the bare states in the Bell basis, we proceed to find its density matrix $\rho_{AB}(t)$ of the D-JCM to  calculate the collapse and revival at each local sites by using the formula,
\begin{subequations} \label{twelve}
\begin{eqnarray}
\langle W_{A}(t)\rangle = Tr[\rho_{A}(t)\sigma_{z}], \\
\langle W_{B}(t)\rangle = Tr[\rho_{B}(t)\sigma_{z}],
\end{eqnarray}
\end{subequations}
\noindent
where $\rho_{i}(t)$($=Tr_j[\rho_{AB}(t)]$) be the reduced density matrix of the subsystem at local site $i,j =A, B$ and $i\neq j$, respectively.

\section{\label{2a} Dressed state solution of Double Jaynes-Cummings model}

For Scenario-I, to solve the D-JCM using the dressed state formalism, we define the initial dressed wave function by a orthogonal transformation of the bare state in Eq.(9) \cite{nath2008,sen2014},
\begin{equation}\label{thirteen}
\left(
  \begin{array}{c}
   \mid 1;{n_A,n_B}\rangle\\
   \mid 2;{n_A,n_B}\rangle\\
   \mid 3;{n_A,n_B}\rangle\\
   \mid 4;{n_A,n_B}\rangle\\
  \end{array}
\right)=T_{\alpha}\left(
  \begin{array}{c}
   \mid {n_A+1,n_B+1}\rangle\\
   \mid {n_A+1,n_B}\rangle\\
   \mid {n_A,n_B+1}\rangle\\
   \mid {n_A,n_B}\rangle\\
  \end{array}
\right).
\end{equation}
In Eq.(13), the transformation matrix $T_{\alpha}$ must satisfy the following consistency condition,
\begin{equation} \label{fourteen}
T_{\alpha}H_{n_A,n_B}^I T_{\alpha}^T=diag(E^{1}_{n_A,n_B}, E^{2}_{n_A,n_B}, E^{3}_{n_A,n_B}, E^{4}_{n_A,n_B}),
\end{equation}
where $H_{n_A,n_B}^I$ and $E_{n_A,n_B}^{i}$ be the interaction Hamiltonian and their eigenvalues in the dressed state basis which are to be evaluated. In Eq.(14), the orthogonal matrix $T_{\alpha}$ is given by,
\begin{equation}\label{fifteen}
T_{\alpha}=\left(
  \begin{array}{cccc}
    \alpha_{11}  & \alpha_{12} & \alpha_{13} & \alpha_{14} \\
    \alpha_{21}  & \alpha_{22} & \alpha_{23} & \alpha_{24} \\
    \alpha_{31}  & \alpha_{32} & \alpha_{33} & \alpha_{34} \\
    \alpha_{41}  & \alpha_{42} & \alpha_{43} & \alpha_{44}
  \end{array}
\right),
\end{equation}
with $\alpha_{ij}$ is parameterized as \cite{bose1980}, 
\begin{eqnarray}
\begin{array}{l}\label{sixteen}
{\alpha _{11}} = {c_1}{c_5} + {s_1}{s_3}{s_4}{s_5}\\
{\alpha _{12}} = {c_1}{s_5}{s_6} + {s_1}{c_3}{c_6} + {s_1}{s_3}{s_4}{c_5}{s_6}\\
{\alpha _{13}} = {s_1}{s_3}{c_4}\\
{\alpha _{14}} =  - {c_1}{s_5}{s_6} - {s_1}{c_3}{s_6} + {s_1}{s_3}{s_4}{c_5}{s_6}\\
{\alpha _{21}} =  - {s_1}{c_2}{c_5} + ({c_1}{c_2}{s_3} - {s_2}{c_3}){s_4}{s_5}\\
{\alpha _{22}} = {s_1}{c_2}{s_5}{s_6} + {c_1}{c_2}{c_3} + {s_2}{s_3}){c_6} + ({c_1}{c_2}{s_3} \\
\qquad - {s_2}{c_3}){s_4}{c_5}{s_6}\\
{\alpha _{23}} = ({c_1}{c_2}{s_3} - {s_2}{c_3}){c_4}\\
{\alpha _{24}} = {s_1}{c_2}{s_5}{c_6} - ({c_1}{c_2}{c_3} + {s_2}{s_3}){c_6} + ({c_1}{c_2}{s_3} \\
\qquad - {s_2}{c_3}){s_4}{c_5}{s_6}\\
{\alpha _{31}} =  - {s_1}{s_2}{c_5} + ({c_1}{s_2}{s_3} + {c_2}{c_3}){s_4}\\
{\alpha _{32}} = {s_1}{s_2}{s_5}{s_6} + ({c_1}{c_2}{c_3} - {c_2}{s_3}){c_6} + ({c_1}{s_2}{s_3} \\
\qquad + {c_2}{c_3}){s_4}{c_5}{s_6}\\
{\alpha _{33}} = ({c_1}{s_2}{s_3} + {c_2}{c_3}){c_4}\\
{\alpha _{34}} = {s_1}{s_2}{s_5}{c_6} - ({c_1}{s_2}{c_3} - {c_2}{s_3}){s_6} + ({c_1}{s_2}{s_3} \\
 \qquad + {c_2}{c_3}){s_4}{c_5}{s_6}\\
{\alpha _{41}} = {c_4}{s_5}\\
{\alpha _{42}} = {c_4}{c_5}{s_6}\\
{\alpha _{43}} =  - {s_4}\\
{\alpha _{44}} = {c_4}{c_5}{c_6}
\end{array}
\end{eqnarray}
with ${s_k} = \sin{\theta _k}$  and ${c_k} = \cos{\theta_k}$ $(k=1,2,3,4,5,6)$. Using the algebra, namely,
\begin{eqnarray}\label{seventeen}
a_A\sigma_A^+\mid 0_A0_B;{n_A+1,n_B+1}\rangle &=&\sqrt{n_A+1}\mid 1_A0_B;{n_A,n_B+1}\rangle, \nonumber \\
a_A\sigma_A^+\mid 0_A1_B;{n_A+1,n_B}\rangle &=&\sqrt{n_A+1}\mid 1_A1_B;{n_A,n_B}\rangle, \nonumber \\
a_B\sigma_B^+\mid 0_A0_B;{n_A+1,n_B+1}\rangle &=&\sqrt{n_B+1}\sigma_B^+\mid 0_A1_B;{n_A+1,n_B}\rangle \nonumber \\
a_B\sigma_B^+\mid 1_A0_B;{n_A,n_B+1}\rangle &=&\sqrt{n_B+1}\mid 1_A1_B;{n_A,n_B}\rangle, \\
{a_A}^\dag\sigma_A^-\mid 1_A0_B;{n_A,n_B+1}\rangle &=&\sqrt{n_A+1}\mid 0_A0_B;{n_A+1,n_B+1}\rangle,  \nonumber \\
{a_A}^\dag\sigma_A^-\mid 1_A1_B;{n_A,n_B}\rangle &=&\sqrt{n_A+1}\mid 0_A1_B;{n_A+1,n_B}\rangle, \nonumber \\
{a_B}^\dag\sigma_B^-\mid 0_A1_B;{n_A+1,n_B}\rangle &=&\sqrt{n_B+1}\mid 0_A0_B;{n_A+1,n_B+1}\rangle,\nonumber \\
{a_B}^\dag\sigma_B^-\mid 1_A1_B;{n_A,n_B}\rangle &=&\sqrt{n_B+1}\mid 1_A0_B;{n_A,n_B+1}\rangle, \nonumber
\end{eqnarray}
it is straightforward to find that the mixing angles to be
\begin{equation}
\begin{split} \label{eighteen}
\theta_1&=\arccos{\frac{1}{\sqrt{6}}}, \quad \theta_2 = \arccos{\frac{2}{\sqrt{5}}}, \quad \theta_3 = \arccos{-\sqrt{\frac{3}{5}}}, \\
\quad \theta_4 &= \arccos{-\frac{\sqrt{3}}{2}}, \quad \theta_5=\arccos{\sqrt{\frac{2}{3}}}, \quad \theta_6 = \arccos{\frac{1}{\sqrt{2}}},
\end{split}
\end{equation}
and in the dressed basis, the interaction Hamiltonian Eq.(3) becomes,
\begin{equation}\label{nineteen}
H_{n_A,n_B}^I=\left(
  \begin{array}{cccc}
    0  & \frac{\Omega_{B}}{2} & \frac{\Omega_{A}}{2} & 0 \\
    \frac{\Omega_{B}}{2} & 0 & 0 & \frac{\Omega_{A}}{2} \\
    \frac{\Omega_{A}}{2} & 0 & 0 & \frac{\Omega_{B}}{2} \\
    0 & \frac{\Omega_{A}}{2} & \frac{\Omega_{B}}{2} & 0 \\
  \end{array}
\right),
\end{equation}
where, $\Omega_{i}=g_i\sqrt{n_i+1}$. The corresponding time-dependent amplitude is found to be,
\begin{equation}\label{twenty}
\left(
  \begin{array}{c}
  c_{01}(t)\\
   c_{11}(t)\\
   c_{11}(t)\\
   c_{01}(t)
  \end{array}
\right)=
 T_{\alpha}^T\left(
  \begin{array}{cccc}
   e^{-iE^{1}_{n_A,n_B}t} & 0 & 0 & 0\\
  0 & e^{-iE^{2}_{n_A,n_B}t} & 0 & 0\\
  0 & 0 & e^{-iE^{3}_{n_A,n_B}t} & 0\\
  0 & 0 & 0 & e^{-iE^{4}_{n_A,n_B}t}
  \end{array}
\right)T_{\alpha}\left(
  \begin{array}{c}
  c_{00}(0)\\
   c_{01}(0)\\
   c_{01}(0)\\
   c_{00}(0)
  \end{array}
\right),
\end{equation}
where, $E^{1}_{n_A,n_B}=-E^{2}_{n_A,n_B}=-\frac{1}{2}(\Omega_{n_A}-\Omega_{n_B})$, and $E^{3}_{n_A,n_B}=-E^{4}_{n_A,n_B}=-\frac{1}{2}(\Omega_{n_A}+\Omega_{n_B})$, respectively. Thus the time-dependent wave function in the Bell basis is found to be,
\begin{eqnarray}\label{twentyone}
\mid \psi_{AB}(t) \rangle &=\sum\limits_{n_A,n_B=0}^{\infty} \{\bar{c}_{00}(t)\mid \Phi^+_{AB};n_A,n_B \rangle +\bar{c}_{01}(t)\mid \Psi^+_{AB};n_A,n_B\rangle \}, 
\end{eqnarray}
where the amplitudes are given by,
\begin{subequations}
\begin{eqnarray}\label{twentytwo}
\bar{c}_{00}(t)&=&c_{00}(0)\cos(\Omega_A+\Omega_B) t-ic_{01}(0)\cos(\Omega_A+\Omega_B) t\\
\bar{c}_{01}(t)&=&c_{01}(0)\cos(\Omega_A+\Omega_B) t-ic_{00}(0)\cos(\Omega_A+\Omega_B) t,
\end{eqnarray}
\end{subequations}
respectively. The corresponding density matrix for Scenario-I is found to be,
\begin{subequations}
\begin{eqnarray}\label{twentythree}
\rho_{11}^{AB}(t)&=&\rho_{44}^{AB}(t)=\rho_{14}^{AB}(t)={\rho_{41}^{AB}}^*(t)=\nonumber \\
& &c_{00}^2(0)\cos^2{\bigg [\frac{1}{2}(\Omega_A+\Omega_B) t}\bigg ]+c_{01}^2(0)\sin^2\bigg [{\frac{1}{2}(\Omega_A+\Omega_B) t}\bigg ], \\
\rho_{22}^{AB}(t)&=& \rho_{33}^{AB}(t)=\rho_{23}^{AB}(t)={\rho_{32}^{AB}}^*(t)=\nonumber \\
& & c_{01}^2(0)\cos^2\bigg [{\frac{1}{2}(\Omega_A+\Omega_B) t}\bigg ]+c_{00}^2(0)\sin^2\bigg [{\frac{1}{2}(\Omega_A+\Omega_B) t}\bigg ], \\
\rho_{12}^{AB}(t)&=&{\rho_{21}^{AB}}^*(t)=\rho_{13}^{AB}(t)={\rho_{31}^{AB}}^*(t)=\nonumber \\
& &c_{00}(0)c_{01}(0)-\frac{i}{2}(c_{00}^2(0)-c_{01}^2(0))\sin{(\Omega_A+\Omega_B) t}.
\end{eqnarray}
\end{subequations}
In the dressed basis, corresponding diagonal part of the Hamiltonian from Eq.(2) is given by,
\begin{equation}\label{twentyfour}
H^0_{n_A,n_B}=\left(
  \begin{array}{cccc}
    H^{11}_{n_A,n_B}  & 0 & 0 & 0 \\
    0 & H^{22}_{n_A,n_B} & 0 & 0 \\
    0 & 0 & H^{33}_{n_A,n_B} & 0 \\
    0 & 0 & 0 & H^{44}_{n_A,n_B}
  \end{array}
\right),
\end{equation}
where the elements are,
\begin{equation}\label{twentyfive}
\begin{split}
H^{11}_{n_A,n_B} &= n_A\omega_{A}+n_B\omega_{B}+(1+n_A)\omega_{A}+(1+n_B)\omega_{B}, \\
H^{22}_{n_A,n_B} &= (1+n_A)\omega_{A}+n_B\omega_{B},\\
H^{33}_{n_A,n_B} &= n_A\omega_{A}+(1+n_B)\omega_{B}, \\
H^{44}_{n_A,n_B} &= n_A\omega_{A}+n_B\omega_{B}.
\end{split}
\end{equation}
From Eq.(19) and (24), the total Hamiltonian is given by $H_n=H^0_{n_A,n_B}+H^I_{n_A,n_B}$ and
the effective energies of the dressed states obtained by diagonalising it,
\begin{subequations}\label{twentysix}
\begin{eqnarray}
E^{1}_{n_A,n_B} &=& E^{0}_{n_A,n_B}-\frac{1}{2}(\mathcal{R}_{n_A}-\mathcal{R}_{n_B}),\\
E^{2}_{n_A,n_B} &=& E^{0}_{n_A,n_B}+\frac{1}{2}(\mathcal{R}_{n_A}-\mathcal{R}_{n_B}),\\
E^{3}_{n_A,n_B} &=& E^{0}_{n_A,n_B}-\frac{1}{2}(\mathcal{R}_{n_A}+\mathcal{R}_{n_B}),\\
E^{4}_{n_A,n_B} &=& E^{0}_{n_A,n_B}+\frac{1}{2}(\mathcal{R}_{n_A}+\mathcal{R}_{n_B}),
\end{eqnarray}
\end{subequations}
where, $E^{0}_{n_A,n_B} = (n_A+\frac{1}{2})\omega_{A}+(n_B+\frac{1}{2})\omega_{B}$ and  $\mathcal{R}_{n_i}=\sqrt{\Omega_{n_i}+\omega_{i}}$, respectively. The appearance of multiple number of dressed states due to the entanglement-induced dressing has put our treatment of dressed state formalism in a advantageous position compare to other approaches of D-JCM \cite{eberly2007,malinovsky2006,jarvis2010,jinshi2010,bahari2018,yan2014,Sainz2007,
Han2010}.

\section{Numerical analysis:}
\par
Following the preestablished protocol, finally we shall make measurement at each local sites by taking projection of the entangled system given by Eq.(12) using the density matrix obtained in Eq.(23). For completeness,
we first study the vacuum Rabi oscillation ($n_A=n_B \equiv 0$) of the D-JCM at each local sites and then proceed to study the collapse and revival oscillation. For Scenario-I, Fig.2 depicts the time evolution of the population inversion at A and B with $c_{00}(0)=\cos\theta$ and $c_{01}(0)=\sin\theta$, respectively, for different values of $\theta$. From there it is
\begin{figure}[h]
  \centering
\includegraphics[scale=1]{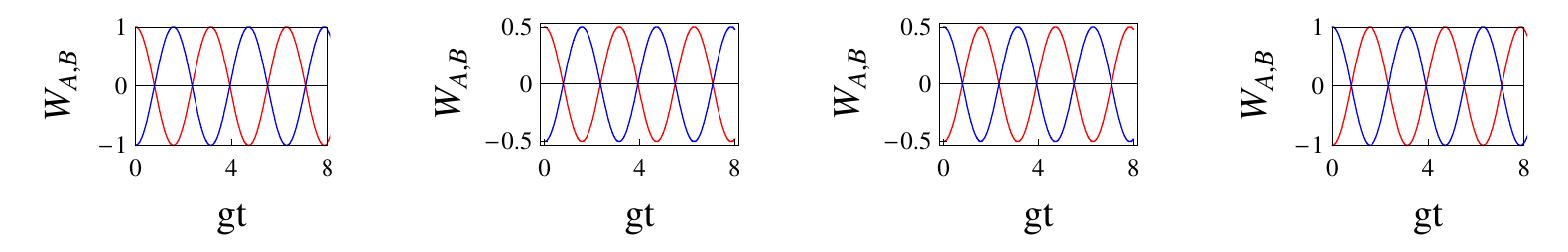}
\begin{flushleft}
FIG.2: Time evolution of the population inversion at site A (Red online) and B (Blue online) for Scenario-I with $\theta=0$, $\theta=\frac{\pi}{6}$, $\theta=\frac{\pi}{3}$ and $\theta=\frac{\pi}{2}$ for $n_A=n_B \equiv 0$ with detuning $\Delta=0$.
\end{flushleft}
\end{figure}
evident that the pattern of Rabi oscillation at Alice and Bob are in the opposite phase to each other. In particular, for $\theta=0$ and $\theta=\frac{\pi}{2}$, which corresponds pure $\Phi^+$ and $\Psi^+$ state respectively, the amplitude of oscillation attains its maximum value which is gradually suppressed with the increase of $\theta$. For $\theta=\frac{\pi}{4}$, which corresponds to the equal admixture of both states, the oscillation completely ceases to exists (not shown).
\begin{figure}[h]
  \centering
\includegraphics[scale=1]{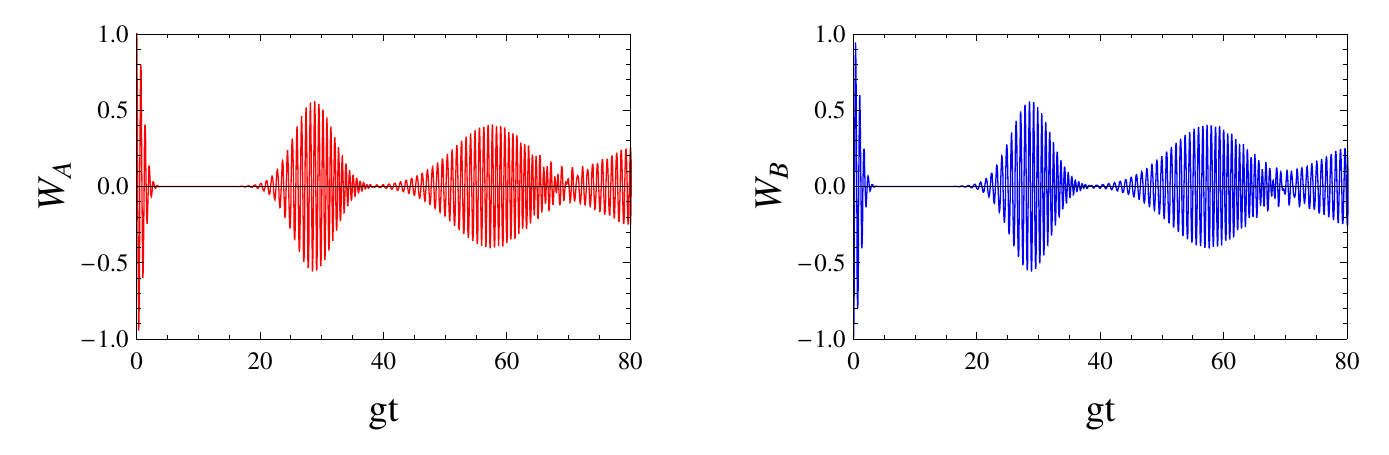}
\begin{flushleft}
FIG.3: Collapse and revival of the population inversion $\langle {W_A(t)}\rangle$ (Red online) and $\langle {W_B(t)}\rangle$ (Blue online) for Case-I with $|\alpha|^2=20$ and $\Delta=0$.
\end{flushleft}
\end{figure}
\par
Finally to study the collapse and revival at local sites, we consider two distinct situations: Case-I: when the state with amplitude $c_{00}(0)$ is initially in the coherent state, i.e., $c_{00}(0)=e^{-|\alpha|^{2}} \frac{\alpha^{n}}{\sqrt{n!}}$ and $c_{01}(0)=0$, and Case-II: when $c_{01}(0)$ is initially in the coherent state, i.e., $c_{00}(0)=0$ and $c_{01}(0) =e^{-|\alpha|^{2}} \frac{\alpha^{n}}{\sqrt{n!}}$, respectively. Fig.3 and 4 illustrate the collapse and revival of the population inversion at Alice and Bob for these two cases
\begin{figure}[h]
  \centering
\includegraphics[scale=1]{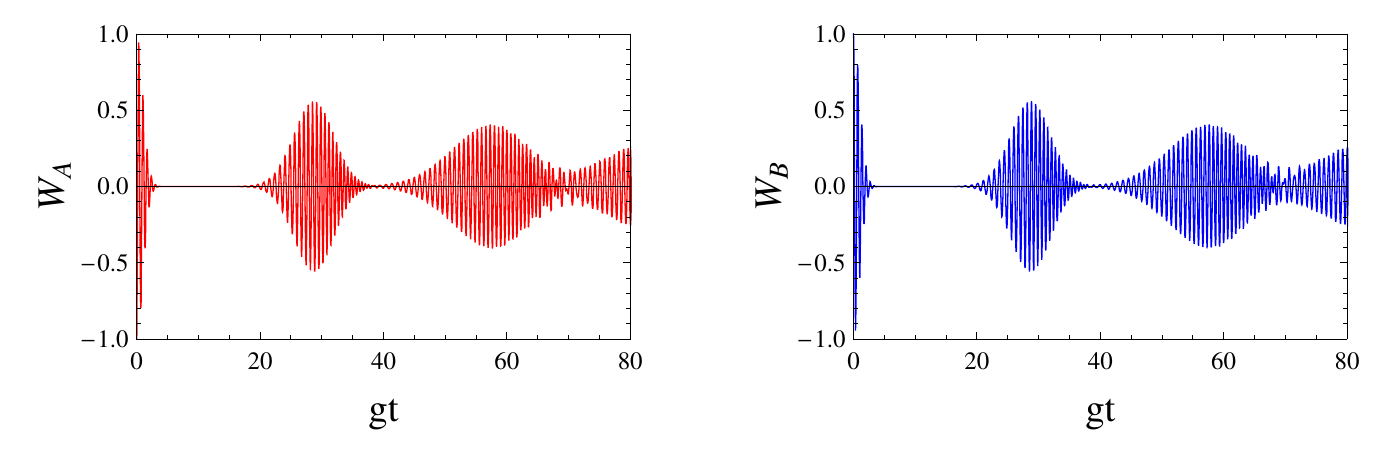}
\begin{flushleft}
FIG.4: Collapse and revival of the Population Inversion $\langle {W_A(t)}\rangle$ and $\langle {W_B(t)}\rangle$ for Case-II with $|\alpha|^2=20$ with $\Delta=0$.
\end{flushleft}
\end{figure}
which are indeed identical to each other. More specifically, the time of collapse and time of revival at local sites are found to be $t_{c}\simeq\frac{\sqrt{2}}{g}$ and $t_{r}\simeq\frac{2\pi |\alpha|}{g}$, which precisely coincide with those of the typical Jaynes-Cummings model \cite{barnett1997}. In other words, it is possible to observe the collapse and revival at `A' and `B' and measure their respective collapse and revival time. The collapse and revival for Scenario-II with other pair of Bell states in Eq.(10) is identical to Scenario-I and therefore we abstain to reiterate it.
\section{Conclusion}
\par
In this paper we have developed a dressed state formalism to solve the lossless Double Jaynes-Cummings model while taking the basis state as the linear superposition of two Bell states. In particular, it is shown that for some specific pairs of Bell states, the model precisely recovers the collapse and revival time of typical JCM at each local sites. Going beyond multifaceted applications of photonic entanglement, the study of the entanglement scenario of atomic or molecular system with two or three levels \cite{sen2012,sen2014} has a very special appeal as they are connected with plethora of quantum-optical phenomena such as, intensity-intensity correlation, squeezing phenomenon, Moller splitting, bunching and anti-bunching effect etc \cite{barnett1997} including the manipulation of the bulk properties of materials such as EIT \cite{sen2015}. Studies of the diverse class of bipartite system within the framework of the D-JCM may provide a handle to explore a new arena of quantum correlation where the manipulation and control of various optical properties could be done nonlocally.
\section{Acknowledgement:}
SS and TKD thank Raman Center for Atomic, Molecular and Optical Sciences (RCAMOS) of Indian Association for the Cultivation of Science, Kolkata, for hospitality.
\bibliography{mainbib}

\begin{thebibliography}{43}%
\makeatletter
\providecommand \@ifxundefined [1]{%
 \@ifx{#1\undefined}
}%
\providecommand \@ifnum [1]{%
 \ifnum #1\expandafter \@firstoftwo
 \else \expandafter \@secondoftwo
 \fi
}%
\providecommand \@ifx [1]{%
 \ifx #1\expandafter \@firstoftwo
 \else \expandafter \@secondoftwo
 \fi
}%
\providecommand \natexlab [1]{#1}%
\providecommand \enquote  [1]{``#1''}%
\providecommand \bibnamefont  [1]{#1}%
\providecommand \bibfnamefont [1]{#1}%
\providecommand \citenamefont [1]{#1}%
\providecommand \href@noop [0]{\@secondoftwo}%
\providecommand \href [0]{\begingroup \@sanitize@url \@href}%
\providecommand \@href[1]{\@@startlink{#1}\@@href}%
\providecommand \@@href[1]{\endgroup#1\@@endlink}%
\providecommand \@sanitize@url [0]{\catcode `\\12\catcode `\$12\catcode
  `\&12\catcode `\#12\catcode `\^12\catcode `\_12\catcode `\%12\relax}%
\providecommand \@@startlink[1]{}%
\providecommand \@@endlink[0]{}%
\providecommand \url  [0]{\begingroup\@sanitize@url \@url }%
\providecommand \@url [1]{\endgroup\@href {#1}{\urlprefix }}%
\providecommand \urlprefix  [0]{URL }%
\providecommand \Eprint [0]{\href }%
\providecommand \doibase [0]{http://dx.doi.org/}%
\providecommand \selectlanguage [0]{\@gobble}%
\providecommand \bibinfo  [0]{\@secondoftwo}%
\providecommand \bibfield  [0]{\@secondoftwo}%
\providecommand \translation [1]{[#1]}%
\providecommand \BibitemOpen [0]{}%
\providecommand \bibitemStop [0]{}%
\providecommand \bibitemNoStop [0]{.\EOS\space}%
\providecommand \EOS [0]{\spacefactor3000\relax}%
\providecommand \BibitemShut  [1]{\csname bibitem#1\endcsname}%
\let\auto@bib@innerbib\@empty
\bibitem [{\citenamefont {Schr{\"{o}}dinger}(1935)}]{schrodinger1935}%
  \BibitemOpen
  \bibfield  {author} {\bibinfo {author} {\bibfnamefont {E.}~\bibnamefont
  {Schr{\"{o}}dinger}},\ }\href@noop {} {\bibfield  {journal} {\bibinfo
  {journal} {Mathematical Proceeding of the Cambridge Philosophical Society}\
  }\textbf {\bibinfo {volume} {31}},\ \bibinfo {pages} {555 } (\bibinfo {year}
  {1935})}\BibitemShut {NoStop}%
\bibitem [{\citenamefont {Schr{\"{o}}dinger}(1936)}]{schrodinger1936}%
  \BibitemOpen
  \bibfield  {author} {\bibinfo {author} {\bibfnamefont {E.}~\bibnamefont
  {Schr{\"{o}}dinger}},\ }\href@noop {} {\bibfield  {journal} {\bibinfo
  {journal} {Mathematical Proceeding of the Cambridge Philosophical Society}\
  }\textbf {\bibinfo {volume} {32}},\ \bibinfo {pages} {446 } (\bibinfo {year}
  {1936})}\BibitemShut {NoStop}%
\bibitem [{\citenamefont {Bennett}\ and\ \citenamefont
  {Brassard}(1984)}]{bb1984}%
  \BibitemOpen
  \bibfield  {author} {\bibinfo {author} {\bibfnamefont {C.~H.}\ \bibnamefont
  {Bennett}}\ and\ \bibinfo {author} {\bibfnamefont {G.}~\bibnamefont
  {Brassard}},\ }\href@noop {} {\bibfield  {journal} {\bibinfo  {journal}
  {Proceedings of IEEE International Conference on Computers}\ }\textbf
  {\bibinfo {volume} {175}},\ \bibinfo {pages} {8} (\bibinfo {year}
  {1984})}\BibitemShut {NoStop}%
\bibitem [{\citenamefont {Rivest}\ \emph {et~al.}(1978)\citenamefont {Rivest},
  \citenamefont {Shamir},\ and\ \citenamefont {Adleman}}]{rsa1978}%
  \BibitemOpen
  \bibfield  {author} {\bibinfo {author} {\bibfnamefont {R.~L.}\ \bibnamefont
  {Rivest}}, \bibinfo {author} {\bibfnamefont {A.}~\bibnamefont {Shamir}}, \
  and\ \bibinfo {author} {\bibfnamefont {L.}~\bibnamefont {Adleman}},\
  }\href@noop {} {\bibfield  {journal} {\bibinfo  {journal} {Communications of
  the ACM}\ }\textbf {\bibinfo {volume} {21}},\ \bibinfo {pages} {120 }
  (\bibinfo {year} {1978})}\BibitemShut {NoStop}%
\bibitem [{\citenamefont {Ekert}(1991)}]{ekert1991}%
  \BibitemOpen
  \bibfield  {author} {\bibinfo {author} {\bibfnamefont {A.~K.}\ \bibnamefont
  {Ekert}},\ }\href@noop {} {\bibfield  {journal} {\bibinfo  {journal} {Phys.
  Rev. Lett.}\ }\textbf {\bibinfo {volume} {67}},\ \bibinfo {pages} {661 }
  (\bibinfo {year} {1991})}\BibitemShut {NoStop}%
\bibitem [{\citenamefont {Nielsen}\ and\ \citenamefont
  {Chuang}(2000)}]{nielsen2000}%
  \BibitemOpen
  \bibfield  {author} {\bibinfo {author} {\bibfnamefont {M.~A.}\ \bibnamefont
  {Nielsen}}\ and\ \bibinfo {author} {\bibfnamefont {I.~L.}\ \bibnamefont
  {Chuang}},\ }\href@noop {} {\emph {\bibinfo {title} {Quantum Information and
  Quantum Computation}}}\ (\bibinfo  {publisher} {Cambridge University Press},\
  \bibinfo {address} {Cambridge},\ \bibinfo {year} {2000})\BibitemShut
  {NoStop}%
\bibitem [{\citenamefont {Aspect}\ \emph {et~al.}(1982)\citenamefont {Aspect},
  \citenamefont {Grangier},\ and\ \citenamefont {Roger}}]{aspect1982}%
  \BibitemOpen
  \bibfield  {author} {\bibinfo {author} {\bibfnamefont {A.}~\bibnamefont
  {Aspect}}, \bibinfo {author} {\bibfnamefont {P.}~\bibnamefont {Grangier}}, \
  and\ \bibinfo {author} {\bibfnamefont {G.}~\bibnamefont {Roger}},\
  }\href@noop {} {\bibfield  {journal} {\bibinfo  {journal} {Phys. Rev. Lett.}\
  }\textbf {\bibinfo {volume} {49}},\ \bibinfo {pages} {91 } (\bibinfo {year}
  {1982})}\BibitemShut {NoStop}%
\bibitem [{\citenamefont {Bell}(1964)}]{bell1964}%
  \BibitemOpen
  \bibfield  {author} {\bibinfo {author} {\bibfnamefont {J.}~\bibnamefont
  {Bell}},\ }\href@noop {} {\bibfield  {journal} {\bibinfo  {journal}
  {Physics}\ }\textbf {\bibinfo {volume} {1}},\ \bibinfo {pages} {195 }
  (\bibinfo {year} {1964})}\BibitemShut {NoStop}%
\bibitem [{\citenamefont {Clauser}\ \emph {et~al.}(1969)\citenamefont
  {Clauser}, \citenamefont {Horne}, \citenamefont {Shimony},\ and\
  \citenamefont {Holt}}]{chsh1969}%
  \BibitemOpen
  \bibfield  {author} {\bibinfo {author} {\bibfnamefont {J.~F.}\ \bibnamefont
  {Clauser}}, \bibinfo {author} {\bibfnamefont {M.~A.}\ \bibnamefont {Horne}},
  \bibinfo {author} {\bibfnamefont {A.}~\bibnamefont {Shimony}}, \ and\
  \bibinfo {author} {\bibfnamefont {R.~A.}\ \bibnamefont {Holt}},\ }\href@noop
  {} {\bibfield  {journal} {\bibinfo  {journal} {Phys. Rev. Lett.}\ }\textbf
  {\bibinfo {volume} {23}},\ \bibinfo {pages} {880} (\bibinfo {year}
  {1969})}\BibitemShut {NoStop}%
\bibitem [{\citenamefont {Einstein}\ \emph {et~al.}(1935)\citenamefont
  {Einstein}, \citenamefont {Podolsky},\ and\ \citenamefont {Rosen}}]{epr1935}%
  \BibitemOpen
  \bibfield  {author} {\bibinfo {author} {\bibfnamefont {A.}~\bibnamefont
  {Einstein}}, \bibinfo {author} {\bibfnamefont {B.}~\bibnamefont {Podolsky}},
  \ and\ \bibinfo {author} {\bibfnamefont {N.}~\bibnamefont {Rosen}},\
  }\href@noop {} {\bibfield  {journal} {\bibinfo  {journal} {Phys. Rev.}\
  }\textbf {\bibinfo {volume} {47}},\ \bibinfo {pages} {777} (\bibinfo {year}
  {1935})}\BibitemShut {NoStop}%
\bibitem [{\citenamefont {Hensen}\ \emph {et~al.}(2015)\citenamefont {Hensen}
  \emph {et~al.}}]{hensen2015}%
  \BibitemOpen
  \bibfield  {author} {\bibinfo {author} {\bibfnamefont {B.}~\bibnamefont
  {Hensen}} \emph {et~al.},\ }\href@noop {} {\bibfield  {journal} {\bibinfo
  {journal} {Phys. Rev. Lett.}\ }\textbf {\bibinfo {volume} {526}},\ \bibinfo
  {pages} {682 } (\bibinfo {year} {2015})}\BibitemShut {NoStop}%
\bibitem [{\citenamefont {Shalm}\ \emph {et~al.}(2015)\citenamefont {Shalm}
  \emph {et~al.}}]{shalm2015}%
  \BibitemOpen
  \bibfield  {author} {\bibinfo {author} {\bibfnamefont {L.~K.}\ \bibnamefont
  {Shalm}} \emph {et~al.},\ }\href@noop {} {\bibfield  {journal} {\bibinfo
  {journal} {Phys. Rev. Lett.}\ }\textbf {\bibinfo {volume} {115}},\ \bibinfo
  {pages} {250402} (\bibinfo {year} {2015})}\BibitemShut {NoStop}%
\bibitem [{\citenamefont {Hill}\ and\ \citenamefont
  {Wootters}(1997)}]{wootters1997}%
  \BibitemOpen
  \bibfield  {author} {\bibinfo {author} {\bibfnamefont {S.}~\bibnamefont
  {Hill}}\ and\ \bibinfo {author} {\bibfnamefont {W.~K.}\ \bibnamefont
  {Wootters}},\ }\href@noop {} {\bibfield  {journal} {\bibinfo  {journal}
  {Phys. Rev. Lett.}\ }\textbf {\bibinfo {volume} {78}},\ \bibinfo {pages}
  {5022} (\bibinfo {year} {1997})}\BibitemShut {NoStop}%
\bibitem [{\citenamefont {Wootters}(1998)}]{wootters1998}%
  \BibitemOpen
  \bibfield  {author} {\bibinfo {author} {\bibfnamefont {W.~K.}\ \bibnamefont
  {Wootters}},\ }\href@noop {} {\bibfield  {journal} {\bibinfo  {journal}
  {Phys. Rev. Lett.}\ }\textbf {\bibinfo {volume} {80}},\ \bibinfo {pages}
  {2245} (\bibinfo {year} {1998})}\BibitemShut {NoStop}%
\bibitem [{\citenamefont {Wootters}(2006)}]{wootters2006}%
  \BibitemOpen
  \bibfield  {author} {\bibinfo {author} {\bibfnamefont {W.~K.}\ \bibnamefont
  {Wootters}},\ }\href@noop {} {\bibfield  {journal} {\bibinfo  {journal} {Int.
  J. Quant. Inf.}\ }\textbf {\bibinfo {volume} {4}},\ \bibinfo {pages} {219}
  (\bibinfo {year} {2006})}\BibitemShut {NoStop}%
\bibitem [{\citenamefont {Rungta}\ \emph {et~al.}(2001)\citenamefont {Rungta},
  \emph {et~al.}}]{rungta2001}%
  \BibitemOpen
  \bibfield  {author} {\bibinfo {author} {\bibfnamefont {P.}~\bibnamefont
  {Rungta}}, ,  \emph {et~al.},\ }\href@noop {} {\bibfield  {journal} {\bibinfo
   {journal} {Phys. Rev. A}\ }\textbf {\bibinfo {volume} {64}},\ \bibinfo
  {pages} {042315} (\bibinfo {year} {2001})}\BibitemShut {NoStop}%
\bibitem [{\citenamefont {Ollivier}\ and\ \citenamefont
  {Zurek}(2001)}]{ollivier2001}%
  \BibitemOpen
  \bibfield  {author} {\bibinfo {author} {\bibfnamefont {H.}~\bibnamefont
  {Ollivier}}\ and\ \bibinfo {author} {\bibfnamefont {W.~H.}\ \bibnamefont
  {Zurek}},\ }\href@noop {} {\bibfield  {journal} {\bibinfo  {journal} {Phys.
  Rev. Lett.}\ }\textbf {\bibinfo {volume} {88}},\ \bibinfo {pages} {017901}
  (\bibinfo {year} {2001})}\BibitemShut {NoStop}%
\bibitem [{\citenamefont {Lewenstein}\ \emph {et~al.}(2001)\citenamefont
  {Lewenstein}, \citenamefont {Kraus}, \citenamefont {Horodecki},\ and\
  \citenamefont {Cirac}}]{lewenstein2001}%
  \BibitemOpen
  \bibfield  {author} {\bibinfo {author} {\bibfnamefont {M.}~\bibnamefont
  {Lewenstein}}, \bibinfo {author} {\bibfnamefont {B.}~\bibnamefont {Kraus}},
  \bibinfo {author} {\bibfnamefont {P.}~\bibnamefont {Horodecki}}, \ and\
  \bibinfo {author} {\bibfnamefont {J.~I.}\ \bibnamefont {Cirac}},\ }\href@noop
  {} {\bibfield  {journal} {\bibinfo  {journal} {Phys. Rev. A}\ }\textbf
  {\bibinfo {volume} {63}},\ \bibinfo {pages} {044304} (\bibinfo {year}
  {2001})}\BibitemShut {NoStop}%
\bibitem [{\citenamefont {G{\"{o}}hne}\ and\ \citenamefont
  {Toth}(2009)}]{guhne2009}%
  \BibitemOpen
  \bibfield  {author} {\bibinfo {author} {\bibfnamefont {O.}~\bibnamefont
  {G{\"{o}}hne}}\ and\ \bibinfo {author} {\bibfnamefont {G.}~\bibnamefont
  {Toth}},\ }\href@noop {} {\bibfield  {journal} {\bibinfo  {journal} {Phys.
  Rep.}\ }\textbf {\bibinfo {volume} {474}},\ \bibinfo {pages} {1 } (\bibinfo
  {year} {2009})}\BibitemShut {NoStop}%
\bibitem [{\citenamefont {Y{\"{o}}nac}\ \emph {et~al.}(2006)\citenamefont
  {Y{\"{o}}nac}, \citenamefont {Yu},\ and\ \citenamefont
  {Eberly}}]{eberly2006}%
  \BibitemOpen
  \bibfield  {author} {\bibinfo {author} {\bibfnamefont {M.}~\bibnamefont
  {Y{\"{o}}nac}}, \bibinfo {author} {\bibfnamefont {T.}~\bibnamefont {Yu}}, \
  and\ \bibinfo {author} {\bibfnamefont {J.}~\bibnamefont {Eberly}},\
  }\href@noop {} {\bibfield  {journal} {\bibinfo  {journal} {J. Phys. B: At.
  Mol. Opt. Phys.}\ }\textbf {\bibinfo {volume} {39}},\ \bibinfo {pages} {S621}
  (\bibinfo {year} {2006})}\BibitemShut {NoStop}%
\bibitem [{\citenamefont {Y{\"{o}}nac}\ \emph {et~al.}(2007)\citenamefont
  {Y{\"{o}}nac}, \citenamefont {Yu},\ and\ \citenamefont
  {Eberly}}]{eberly2007}%
  \BibitemOpen
  \bibfield  {author} {\bibinfo {author} {\bibfnamefont {M.}~\bibnamefont
  {Y{\"{o}}nac}}, \bibinfo {author} {\bibfnamefont {T.}~\bibnamefont {Yu}}, \
  and\ \bibinfo {author} {\bibfnamefont {J.~H.}\ \bibnamefont {Eberly}},\
  }\href@noop {} {\bibfield  {journal} {\bibinfo  {journal} {J. Phys. B: At.
  Mol. Opt. Phys.}\ }\textbf {\bibinfo {volume} {40}},\ \bibinfo {pages} {S45}
  (\bibinfo {year} {2007})}\BibitemShut {NoStop}%
\bibitem [{\citenamefont {Yu}\ and\ \citenamefont {Eberly}(2009)}]{eberly2009}%
  \BibitemOpen
  \bibfield  {author} {\bibinfo {author} {\bibfnamefont {T.}~\bibnamefont
  {Yu}}\ and\ \bibinfo {author} {\bibfnamefont {J.~H.}\ \bibnamefont
  {Eberly}},\ }\href@noop {} {\bibfield  {journal} {\bibinfo  {journal}
  {Science}\ }\textbf {\bibinfo {volume} {30}},\ \bibinfo {pages} {598 }
  (\bibinfo {year} {2009})}\BibitemShut {NoStop}%
\bibitem [{\citenamefont {Y{\"{o}}nac}\ and\ \citenamefont
  {Eberly}(2008)}]{eberly2008}%
  \BibitemOpen
  \bibfield  {author} {\bibinfo {author} {\bibfnamefont {M.}~\bibnamefont
  {Y{\"{o}}nac}}\ and\ \bibinfo {author} {\bibfnamefont {J.~H.}\ \bibnamefont
  {Eberly}},\ }\href@noop {} {\bibfield  {journal} {\bibinfo  {journal} {Opt.
  Lett.}\ }\textbf {\bibinfo {volume} {33}},\ \bibinfo {pages} {270} (\bibinfo
  {year} {2008})}\BibitemShut {NoStop}%
\bibitem [{\citenamefont {Vladimir S.~Malinovsky}\ and\ \citenamefont
  {Ignacio}(2006)}]{malinovsky2006}%
  \BibitemOpen
  \bibfield  {author} {\bibinfo {author} {\bibfnamefont {V.~S.}\ \bibnamefont
  {Vladimir S.~Malinovsky}}\ and\ \bibinfo {author} {\bibfnamefont {R.~S.}\
  \bibnamefont {Ignacio}},\ }\href@noop {} {\bibfield  {journal} {\bibinfo
  {journal} {Phys. Rev. Lett.}\ }\textbf {\bibinfo {volume} {96}},\ \bibinfo
  {pages} {050502:1} (\bibinfo {year} {2006})}\BibitemShut {NoStop}%
\bibitem [{\citenamefont {Saha}\ \emph {et~al.}(2010)\citenamefont {Saha},
  \citenamefont {Majumder}, \citenamefont {Singh},\ and\ \citenamefont
  {Nayak}}]{saha2010}%
  \BibitemOpen
  \bibfield  {author} {\bibinfo {author} {\bibfnamefont {P.}~\bibnamefont
  {Saha}}, \bibinfo {author} {\bibfnamefont {A.}~\bibnamefont {Majumder}},
  \bibinfo {author} {\bibfnamefont {S.}~\bibnamefont {Singh}}, \ and\ \bibinfo
  {author} {\bibfnamefont {N.}~\bibnamefont {Nayak}},\ }\href@noop {}
  {\bibfield  {journal} {\bibinfo  {journal} {Int. J. Quant. Inform.}\ }\textbf
  {\bibinfo {volume} {8}},\ \bibinfo {pages} {1397} (\bibinfo {year}
  {2010})}\BibitemShut {NoStop}%
\bibitem [{\citenamefont {Jarvis}\ \emph {et~al.}(2010)\citenamefont {Jarvis}
  \emph {et~al.}}]{jarvis2010}%
  \BibitemOpen
  \bibfield  {author} {\bibinfo {author} {\bibfnamefont {C.~E.~A.}\
  \bibnamefont {Jarvis}} \emph {et~al.},\ }\href@noop {} {\bibfield  {journal}
  {\bibinfo  {journal} {J. Opt. Soc. Am. B}\ }\textbf {\bibinfo {volume}
  {27}},\ \bibinfo {pages} {A164} (\bibinfo {year} {2010})}\BibitemShut
  {NoStop}%
\bibitem [{\citenamefont {Jin-Shi}\ \emph {et~al.}(2010)\citenamefont {Jin-Shi}
  \emph {et~al.}}]{jinshi2010}%
  \BibitemOpen
  \bibfield  {author} {\bibinfo {author} {\bibfnamefont {X.}~\bibnamefont
  {Jin-Shi}} \emph {et~al.},\ }\href@noop {} {\bibfield  {journal} {\bibinfo
  {journal} {Phys. Rev. Lett.}\ }\textbf {\bibinfo {volume} {104}},\ \bibinfo
  {pages} {100502:1} (\bibinfo {year} {2010})}\BibitemShut {NoStop}%
\bibitem [{\citenamefont {Bahari}\ \emph {et~al.}(2018)\citenamefont {Bahari},
  \citenamefont {Spiller}, \citenamefont {Dooley}, \citenamefont {Hayes},\ and\
  \citenamefont {McCrossan}}]{bahari2018}%
  \BibitemOpen
  \bibfield  {author} {\bibinfo {author} {\bibfnamefont {I.}~\bibnamefont
  {Bahari}}, \bibinfo {author} {\bibfnamefont {T.~P.}\ \bibnamefont {Spiller}},
  \bibinfo {author} {\bibfnamefont {S.}~\bibnamefont {Dooley}}, \bibinfo
  {author} {\bibfnamefont {A.}~\bibnamefont {Hayes}}, \ and\ \bibinfo {author}
  {\bibfnamefont {F.}~\bibnamefont {McCrossan}},\ }\href@noop {} {\bibfield
  {journal} {\bibinfo  {journal} {Int. J. Quant. Inform.}\ }\textbf {\bibinfo
  {volume} {16}},\ \bibinfo {pages} {1850017} (\bibinfo {year}
  {2018})}\BibitemShut {NoStop}%
\bibitem [{\citenamefont {Yan}\ and\ \citenamefont {Zhang}(2014)}]{yan2014}%
  \BibitemOpen
  \bibfield  {author} {\bibinfo {author} {\bibfnamefont {X.-Q.}\ \bibnamefont
  {Yan}}\ and\ \bibinfo {author} {\bibfnamefont {B.-Y.}\ \bibnamefont
  {Zhang}},\ }\href@noop {} {\bibfield  {journal} {\bibinfo  {journal} {Ann.
  Phys.}\ }\textbf {\bibinfo {volume} {349}},\ \bibinfo {pages} {350} (\bibinfo
  {year} {2014})}\BibitemShut {NoStop}%
\bibitem [{\citenamefont {Sainz}\ and\ \citenamefont
  {Bj\"ork}(2007)}]{Sainz2007}%
  \BibitemOpen
  \bibfield  {author} {\bibinfo {author} {\bibfnamefont {I.}~\bibnamefont
  {Sainz}}\ and\ \bibinfo {author} {\bibfnamefont {G.}~\bibnamefont
  {Bj\"ork}},\ }\href {\doibase 10.1103/PhysRevA.76.042313} {\bibfield
  {journal} {\bibinfo  {journal} {Phys. Rev. A}\ }\textbf {\bibinfo {volume}
  {76}},\ \bibinfo {pages} {042313} (\bibinfo {year} {2007})}\BibitemShut
  {NoStop}%
\bibitem [{\citenamefont {Han}(2010)}]{Han2010}%
  \BibitemOpen
  \bibfield  {author} {\bibinfo {author} {\bibfnamefont {F.}~\bibnamefont
  {Han}},\ }\href {\doibase 10.1007/s11434-010-3149-9} {\bibfield  {journal}
  {\bibinfo  {journal} {Chinese Science Bulletin}\ }\textbf {\bibinfo {volume}
  {55}},\ \bibinfo {pages} {1758} (\bibinfo {year} {2010})}\BibitemShut
  {NoStop}%
\bibitem [{\citenamefont {Joshi}\ and\ \citenamefont {Xiao}(2003)}]{joshi2003}%
  \BibitemOpen
  \bibfield  {author} {\bibinfo {author} {\bibfnamefont {A.}~\bibnamefont
  {Joshi}}\ and\ \bibinfo {author} {\bibfnamefont {M.}~\bibnamefont {Xiao}},\
  }\href@noop {} {\bibfield  {journal} {\bibinfo  {journal} {Phys. Lett. A}\
  }\textbf {\bibinfo {volume} {317}},\ \bibinfo {pages} {370 } (\bibinfo {year}
  {2003})}\BibitemShut {NoStop}%
\bibitem [{\citenamefont {Paspalakis}\ and\ \citenamefont
  {Knight}(1999)}]{paspalakis1999}%
  \BibitemOpen
  \bibfield  {author} {\bibinfo {author} {\bibfnamefont {E.}~\bibnamefont
  {Paspalakis}}\ and\ \bibinfo {author} {\bibfnamefont {P.~L.}\ \bibnamefont
  {Knight}},\ }\href@noop {} {\bibfield  {journal} {\bibinfo  {journal} {J.
  Phys. B: At. Mol. Opt. Phys.}\ }\textbf {\bibinfo {volume} {4}},\ \bibinfo
  {pages} {S372} (\bibinfo {year} {1999})}\BibitemShut {NoStop}%
\bibitem [{\citenamefont {Paspalakis}\ \emph {et~al.}(2002)\citenamefont
  {Paspalakis}, \citenamefont {Kylstra},\ and\ \citenamefont
  {Knight}}]{paspalakis2002}%
  \BibitemOpen
  \bibfield  {author} {\bibinfo {author} {\bibfnamefont {E.}~\bibnamefont
  {Paspalakis}}, \bibinfo {author} {\bibfnamefont {N.~J.}\ \bibnamefont
  {Kylstra}}, \ and\ \bibinfo {author} {\bibfnamefont {P.~L.}\ \bibnamefont
  {Knight}},\ }\href@noop {} {\bibfield  {journal} {\bibinfo  {journal} {Phys.
  Rev. A}\ }\textbf {\bibinfo {volume} {65}},\ \bibinfo {pages} {053808}
  (\bibinfo {year} {2002})}\BibitemShut {NoStop}%
\bibitem [{\citenamefont {Ham}\ and\ \citenamefont {Hemmer}(2000)}]{ham2000}%
  \BibitemOpen
  \bibfield  {author} {\bibinfo {author} {\bibfnamefont {B.~S.}\ \bibnamefont
  {Ham}}\ and\ \bibinfo {author} {\bibfnamefont {P.~R.}\ \bibnamefont
  {Hemmer}},\ }\href@noop {} {\bibfield  {journal} {\bibinfo  {journal} {Phys.
  Rev. Lett.}\ }\textbf {\bibinfo {volume} {84}},\ \bibinfo {pages} {4080}
  (\bibinfo {year} {2000})}\BibitemShut {NoStop}%
\bibitem [{\citenamefont {Harris}\ and\ \citenamefont
  {Yamamoto}(1998)}]{harris1998}%
  \BibitemOpen
  \bibfield  {author} {\bibinfo {author} {\bibfnamefont {S.~E.}\ \bibnamefont
  {Harris}}\ and\ \bibinfo {author} {\bibfnamefont {Y.}~\bibnamefont
  {Yamamoto}},\ }\href@noop {} {\bibfield  {journal} {\bibinfo  {journal}
  {Phys. Rev. Lett.}\ }\textbf {\bibinfo {volume} {81}},\ \bibinfo {pages}
  {3611} (\bibinfo {year} {1998})}\BibitemShut {NoStop}%
\bibitem [{\citenamefont {Nath}\ \emph {et~al.}(2008)\citenamefont {Nath},
  \citenamefont {Dey}, \citenamefont {Sen},\ and\ \citenamefont
  {Gangopadhyay}}]{nath2008}%
  \BibitemOpen
  \bibfield  {author} {\bibinfo {author} {\bibfnamefont {M.~R.}\ \bibnamefont
  {Nath}}, \bibinfo {author} {\bibfnamefont {T.~K.}\ \bibnamefont {Dey}},
  \bibinfo {author} {\bibfnamefont {S.}~\bibnamefont {Sen}}, \ and\ \bibinfo
  {author} {\bibfnamefont {G.}~\bibnamefont {Gangopadhyay}},\ }\href@noop {}
  {\bibfield  {journal} {\bibinfo  {journal} {Pramana - J. Phys}\ }\textbf
  {\bibinfo {volume} {77}},\ \bibinfo {pages} {141} (\bibinfo {year}
  {2008})}\BibitemShut {NoStop}%
\bibitem [{\citenamefont {Sen}\ \emph {et~al.}(2012)\citenamefont {Sen},
  \citenamefont {Nath}, \citenamefont {Dey},\ and\ \citenamefont
  {Gangopadhyay}}]{sen2012}%
  \BibitemOpen
  \bibfield  {author} {\bibinfo {author} {\bibfnamefont {S.}~\bibnamefont
  {Sen}}, \bibinfo {author} {\bibfnamefont {M.~R.}\ \bibnamefont {Nath}},
  \bibinfo {author} {\bibfnamefont {T.~K.}\ \bibnamefont {Dey}}, \ and\
  \bibinfo {author} {\bibfnamefont {G.}~\bibnamefont {Gangopadhyay}},\
  }\href@noop {} {\bibfield  {journal} {\bibinfo  {journal} {Ann. Phys.}\
  }\textbf {\bibinfo {volume} {327}},\ \bibinfo {pages} {224–} (\bibinfo
  {year} {2012})}\BibitemShut {NoStop}%
\bibitem [{\citenamefont {Sen}\ and\ \citenamefont {Ahmed}(2014)}]{sen2014}%
  \BibitemOpen
  \bibfield  {author} {\bibinfo {author} {\bibfnamefont {S.}~\bibnamefont
  {Sen}}\ and\ \bibinfo {author} {\bibfnamefont {H.}~\bibnamefont {Ahmed}},\
  }\href@noop {} {\bibfield  {journal} {\bibinfo  {journal} {J. Math. Phys.}\
  }\textbf {\bibinfo {volume} {55}},\ \bibinfo {pages} {122105 : 1} (\bibinfo
  {year} {2014})}\BibitemShut {NoStop}%
\bibitem [{\citenamefont {Jaynes}\ and\ \citenamefont
  {Cummings}(1963)}]{jcm1963}%
  \BibitemOpen
  \bibfield  {author} {\bibinfo {author} {\bibfnamefont {E.~T.}\ \bibnamefont
  {Jaynes}}\ and\ \bibinfo {author} {\bibfnamefont {F.~W.}\ \bibnamefont
  {Cummings}},\ }\href@noop {} {\bibfield  {journal} {\bibinfo  {journal}
  {Proc. IEEE}\ }\textbf {\bibinfo {volume} {51}},\ \bibinfo {pages} {89}
  (\bibinfo {year} {1963})}\BibitemShut {NoStop}%
\bibitem [{\citenamefont {Barnett}\ and\ \citenamefont
  {Radmore}(1997)}]{barnett1997}%
  \BibitemOpen
  \bibfield  {author} {\bibinfo {author} {\bibfnamefont {S.~M.}\ \bibnamefont
  {Barnett}}\ and\ \bibinfo {author} {\bibfnamefont {P.~M.}\ \bibnamefont
  {Radmore}},\ }\href@noop {} {\emph {\bibinfo {title} {Methods in Theoretical
  Quantum Optics}}}\ (\bibinfo  {publisher} {Clarendon Press, Oxford},\
  \bibinfo {address} {Oxford},\ \bibinfo {year} {1997})\ p.~\bibinfo {pages}
  {23}\BibitemShut {NoStop}%
\bibitem [{\citenamefont {Bose}\ and\ \citenamefont {Pascos}(1980)}]{bose1980}%
  \BibitemOpen
  \bibfield  {author} {\bibinfo {author} {\bibfnamefont {S.~K.}\ \bibnamefont
  {Bose}}\ and\ \bibinfo {author} {\bibfnamefont {E.~A.}\ \bibnamefont
  {Pascos}},\ }\href@noop {} {\bibfield  {journal} {\bibinfo  {journal} {Nucl.
  Phys.}\ }\textbf {\bibinfo {volume} {B169}},\ \bibinfo {pages} {384}
  (\bibinfo {year} {1980})}\BibitemShut {NoStop}%
\bibitem [{\citenamefont {Sen}\ \emph {et~al.}(2015)\citenamefont {Sen},
  \citenamefont {Dey}, \citenamefont {Nath},\ and\ \citenamefont
  {Gangopadhyay}}]{sen2015}%
  \BibitemOpen
  \bibfield  {author} {\bibinfo {author} {\bibfnamefont {S.}~\bibnamefont
  {Sen}}, \bibinfo {author} {\bibfnamefont {T.~K.}\ \bibnamefont {Dey}},
  \bibinfo {author} {\bibfnamefont {M.~R.}\ \bibnamefont {Nath}}, \ and\
  \bibinfo {author} {\bibfnamefont {G.}~\bibnamefont {Gangopadhyay}},\
  }\href@noop {} {\bibfield  {journal} {\bibinfo  {journal} {J. Mod. Opt.}\
  }\textbf {\bibinfo {volume} {62}},\ \bibinfo {pages} {166–} (\bibinfo
  {year} {2015})}\BibitemShut {NoStop}%
\end{thebibliography}%
\end{document}